\documentclass[12pt]{article}
\usepackage{graphicx}
\usepackage{latexsym}

\newcommand{\beq}{\begin{equation}}
\newcommand{\eeq}{\end{equation}}
\newcommand{\bea}{\begin{eqnarray}} 
\newcommand{\eea}{\end{eqnarray}}
\newcommand{\vs}[1]{\vspace{#1 mm}}

\renewcommand{\a}{\alpha}
\renewcommand{\b}{\beta}

\renewcommand{\d}{\delta}

\newcommand{\om}{\omega}
\newcommand{\zb}{\bar{z}}

\def\bbox{{\,\lower0.9pt\vbox{\hrule \hbox{\vrule height 0.2 cm
\hskip 0.2 cm \vrule height 0.2 cm}\hrule}\,}}
\newcommand{\dsl}{\pa \kern-0.5em /}
\newcommand{\la}{\lambda}

\newcommand{\pa}{\partial}
\newcommand{\si}{\sigma}
\renewcommand{\t}{\theta}

\newcommand{\ut}{\tilde{U}}
\newcommand{\Z}{\mbox{$Z\hspace{-2mm}Z$}}
\newcommand{\uf}{{\underline f}}
\newcommand{\ul}{{\underline l}}

\newcommand{\nn}{\nonumber\\ }

\begin{document}

\topmargin30pt
\oddsidemargin 0mm
\renewcommand{\thefootnote}{\fnsymbol{footnote}}
\begin{titlepage}

\setcounter{page}{0}

\vs{10}
\begin{center}

{\Large\bf D-branes in asymmetrically gauged WZW models}\\[.2cm]
{\Large\bf and axial-vector duality}
\vs{15}

{\large Mark A. Walton}\footnote{walton@uleth.ca} and 
{\large Jian-Ge Zhou}\footnote{jiange.zhou@uleth.ca}

\vs{10}
{\em Physics Department, University of Lethbridge,
Lethbridge, Alberta, Canada T1K 3M4} \\
\end{center}

\vs{15}
\centerline{{\bf{Abstract}}}
\vs{5}

We construct D-branes in a left-right asymmetrically gauged
WZW model, with the gauge subgroup embedded differently  
on the left and the right of the group element.
The symmetry-preserving boundary conditions for the group-valued field $g$ 
are described, and
the corresponding action is found. When the subgroup $H=U(1)$,  
we can implement T-duality on the axially gauged
WZW action; an orbifold of the vectorially gauged 
theory is produced.
For the parafermion $SU(2)/U(1)$ coset model, a $\sigma$-model 
is obtained with vanishing
gauge field on D-branes. We show that a boundary condition surviving 
from the $SU(2)$ parent theory 
characterizes D-branes in the parafermion theory, 
determining the shape of A-branes. 
The gauge field on B-branes 
is obtained from the boundary condition for A-branes, 
by the orbifold construction and T-duality. These gauge fields  
stabilize the B-branes.


\end{titlepage}
\newpage

\renewcommand{\thefootnote}{\arabic{footnote}}
\setcounter{footnote}{0}

\section{Introduction}

The pioneering 
work of Maldacena, Moore and Seiberg~\cite{MMS} initiated the study 
of the geometry of
D-branes in coset models \cite{KG}-\cite{YH}. 
The bosonic theory they focused on was the parafermion 
$SU(2)/U(1)$ coset model. 
A-branes were studied using rational
conformal field theory (CFT) via the Cardy construction. By 
$\Z_k$-orbifolding and T-duality, 
B-branes were also obtained. 
 
In~\cite{KG}\cite{ES},  it was
shown that A-branes can be given a geometrical interpretation
in the vectorially gauged Wess-Zumino-Witten (WZW) 
model \cite{GK}\cite{KPS}. 
There the boundary value of the
group-valued field $g$  was found to be in a product of two
conjugacy classes -- one of $G$ and the other of $H$. 
This boundary form was justified in \cite{KRWZ}, 
where it was derived from the corresponding gluing conditions. 
The non-commutative gauge theories dictating the dynamics
of  D-branes in $G/H$ coset models were constructed in \cite{FS}.

There is a one-to-one Cardy correspondence between A-type
boundary states and bulk primary fields in the $SU(2)/U(1)$
parafermion theory \cite{MMS}.
For B-branes the Cardy correspondence does not hold, and so 
we have less understanding of B-branes than  A-branes. 
For instance, it is unclear how to decide if B-branes are stable and if they
can be described by some kind of gauged WZW model. 
From the construction of B-branes in \cite{MMS}, 
it seems that B-branes are related to 
the axially gauged WZW model. For open strings, however,
only the vectorially gauged WZW model 
\cite{KG}\cite{ES}\cite{KRWZ} has been treated. It is
interesting therefore to study D-branes in  
left-right asymmetrically gauged WZW models, including axially gauged 
WZW models as  special cases.

When the gauge subgroup is abelian,  
there is an axial-vector duality \cite{DVV}\cite{EK} 
in the coset model for closed strings. 
Considered as $\sigma$-models, the 
axial and vector gauging of an abelian chiral symmetry 
leads to different target spaces; one may even be singular 
when the other is regular \cite{DVV}. 
The corresponding axially and vectorially gauged 
WZW models describe the same coset 
CFT, however.\,\footnote{Exact abelian dualities are 
well understood \cite{EK1}\cite{AABL} in the case of compact
groups. In the noncompact case we know that axial-vector duality
is exact only for abelian cosets possessing appropriate
Weyl symmetries \cite{GEK}.} 
One may wonder
whether there is an  axial-vector duality in
the coset model for open strings.

Here we discuss D-branes in the left-right asymmetrically gauged
WZW models, with different embeddings of the same gauge
subgroup acting on the left and on the right.
We construct the left-right asymmetrically gauged
WZW action for open strings, and find the boundary
condition for group-valued field $g$ which preserves the  
left-right asymmetric symmetry. 
The methods of \cite{KRWZ} make this straightforward. 

When the subgroup $H$
is abelian, we obtain the axially gauged
WZW action. 
We then implement
T-duality to get the vectorially gauged WZW action. 
When we do T-duality, we find there is a 
crucial change of the boundary condition for the $U(1)$ coordinate
$X$ and its dual  $\tilde{X}$.
The angle $X$ that parametrizes the $U(1)$ subgroup takes 
values in $[0,2\pi]$, 
but the range of values of the
dual angle $\tilde{X}$ is instead $[0,\frac{2\pi}{k}]$.
This is because the axially gauged WZW model is T-dual to the 
$\Z_{k}$ orbifold of the vectorially gauged WZW model.

After constructing D-branes in left-right asymmetrically gauged
WZW model, we specialize to the $SU(2)/U(1)$ coset model,
described by a vectorially gauged WZW model.
From the 
WZW action, we obtain a $\si$-model with vanishing
gauge field on D-branes. Since the geometry of the resulting
disk is conformally flat and the gauge field on D-branes
vanishes, the remaining boundary condition for D-branes 
in the $SU(2)$ parent theory carries over to the coset model. 
From this, we see that the shape
of an A-brane in the $SU(2)/U(1)$ coset model is a straight line. 
This observation is supported by scattering amplitudes
between the boundary states for A-branes and the closed
string states \cite{FFFS}\cite{MMS}. In the parafermion theory
there is global $U(1)$ symmetry, but
the consistency of the gauged WZW model for open strings
does not demand that the $U(1)$ parameter be quantized \cite{KRWZ}. If we
insist on the Cardy correspondence, the scattering amplitudes
between the boundary states for A-branes and the closed
string states indicate that the $U(1)$ symmetry has to be
broken to  $\Z_{2k}$ symmetry. That is, only the $2k$ points 
in a single $\Z_{2k}$ orbit 
on the disk boundary are valid as endpoints of A1-branes. 
The selection rule eliminates
half of these endpoints, and we are left with a 
$\Z_k$ symmetry. 

Since the resulting $\si$-model possesses
$\Z_k$ symmetry, we construct its $\Z_k$ orbifold. Then 
we can implement T-duality. We find that 
the original conformally flat disk is mapped to another 
conformally flat disk, and the gauge field on B-branes 
can be obtained from the boundary condition for A-branes.
The resulting $\si$-model yields the boundary
condition for B-branes, showing that their shapes
are centered disks. In addition, we find gauge
field strengths that stabilize the B-branes -- 
according to \cite{MMS},  
they prevent the B-branes from  decaying by shrinking or by 
displacing off the center. 
Finally, the $\si$-model action 
for B-branes can be recast into the axially gauged WZW action. 
This manifests
the axial-vector duality in the $SU(2)/U(1)$ coset model
for open strings. 

The layout of this paper is as follows. In section 2 we 
discuss D-branes in the asymmetrically gauged WZW model, and
construct the corresponding action for open
strings. We also show how axial-vector duality can
be realized in this model.
In section 3 we
consider the vectorially $U(1)$-gauged WZW model for open strings. 
We extract
the boundary conditions for A-branes in the $SU(2)/U(1)$ coset model
for open strings, and compare them with results obtained 
from the scattering amplitudes
between the boundary states for A-branes and the closed
string states. 
We obtain the $\si$-model for B-branes in section 4, by 
constructing the theory T-dual to an orbifold of that 
describing A-branes. Consequently, 
the gauge field on B-branes is calculated  
from the boundary conditions for A-branes. 
Also discussed is the axial-vector duality in the $SU(2)/U(1)$ coset model
for open strings.
In section 5 we present our summary and discussion.

\section{D-branes in asymmetrically gauged WZW models}

To describe our notation, 
we start with the vectorially gauged WZW model for open strings. Let
$G$ be a compact, simply connected, Lie group. The $G/H$
coset CFT, where $H$ is a subgroup of $G$, can be 
described by a gauged WZW action with the vector symmetry
\beq
g\rightarrow vgv^{-1}
\eeq
gauged
away. Here $g\in G$ and $v\in H\subset G$. 

For closed strings, the gauged WZW action is \cite{GK},\cite{KPS}
\beq
 S^{G/H}\ =\ S(U^{-1}g\ut)-S(U^{-1}\ut)\ =\ S({\hat g})-S(h)\ ,
\label{SGHU}
\eeq
with 
\beq
 {\hat g}=U^{-1}g\ut\in G,\ \ \ \ h=U^{-1}\ut\in H,\ \ \ \ g\in G,\ \ \ \ 
   U,\ut\in H
\label{gth}
\eeq
and gauge fields 
\beq
 A_z=\pa_z \ut\ut^{-1},\ \ \ \ A_{\bar{z}}=\pa_{\bar{z}}UU^{-1}\ . 
\label{A}
\eeq
Here $S$ is the WZW action for the group $G$. 
 
The Wess-Zumino term (part of the WZW action) 
is not well-defined for a worldsheet 
$\Sigma$ with a boundary, however. 
The remedy is to introduce an auxiliary disk $D$ for each hole in $\Sigma$
with boundaries common with those of $\Sigma$ \cite{AS}. For simplicity,
we consider the situation with a single hole.
The map $g$ from $\Sigma$ to $G$ is then 
extended to a map from 
the extended worldsheet $\Sigma\cup D$. The disk $D$ is mapped to
the product of two conjugacy classes for $G$ and 
$H$. Then~\cite{KG}\cite{ES}\cite{KRWZ}
\bea
 S^{G/H}&=&\frac{k}{4\pi}\left(\int_\Sigma d^2z\,L^{\rm kin}({\hat g})
  +\int_{B}\chi({\hat g})-\int_D \om ({\hat g})\right)\nn
 &-&\frac{k_H}{4\pi}\left(\int_\Sigma d^2z\,L^{\rm kin}(h)
  +\int_{B}\chi(h)-\int_D \om (h)\right)\ , 
\label{Sopen}
\eea
is the gauged WZW action for open strings. Here 
$B$ is a three-dimensional manifold bounded by $\Sigma\cup D$.
$\om ({\hat g})$ and $\om (h)$ are the 
Alekseev-Schomerus two-forms defined on the conjugacy 
classes of $G$ and $H$, respectively. 

In the form most useful to us, the action 
of the vectorially gauged WZW model for open string 
is~\cite{KG}\cite{ES}\cite{KRWZ} 
\bea
S^{G/H}&=&\frac{k}{4\pi}\left(\int_\Sigma d^2z\,L^{\rm kin}(g)
+\int_B\chi(g)\right)\nn
&+&\frac{k}{2\pi}\int_\Sigma d^2z\,tr\left\{
A_{z}\pa_{\zb}gg^{-1}-A_{\zb}g^{-1}\pa_{z}g+A_z gA_{\zb} g^{-1}-A_z A_{\zb}
\right\}\nn
&-&\frac{k}{4\pi}\int_D \,\left\{
\Omega(n;\uf)+\Omega(p;\ul)+tr(dc_Hc_H^{-1}c_G^{-1}dc_G)\right\}
\label{gh}
\eea
where $B$ is a three-dimensional manifold
bounded by $\Sigma\cup D$, i.e. $\partial B = \Sigma\cup D$. 
$c_G$ and $c_H$ are elements of fixed conjugacy classes
of $G$ and $H$, respectively. More precisely, let $\tau$ parametrize 
the boundary $\partial\Sigma=-\pa D$. We write 
$c_G=n(\tau)\uf n(\tau)^{-1}$ with $\uf,n\in G$. The underline 
of $\uf$ indicates it is a fixed element,  
independent of $\tau$. Similarly, we write 
$c_H=p(\tau)\ul p(\tau)^{-1}$, $\ul,p\in H$, with $\ul$ fixed. 
The two-forms $\Omega$ are defined by
\beq
 \Omega(n;\uf)=tr(n^{-1}dn\uf n^{-1}dn\uf^{-1}),\ \ \ \ 
  \Omega(p;\ul)=tr(p^{-1}dp\ul p^{-1}dp\ul^{-1})\ \ .
\label{2form}
\eeq
Finally, $L^{\rm kin}(g)=tr(\pa_z g\pa_{\zb}g^{-1})$, $\chi(g)=
\frac{1}{3}tr(dgg^{-1})^3$. These latter 
satisfy the Polyakov-Wiegmann identities 
\bea
&&L^{\rm kin}(g_{1}g_{2})=L^{\rm kin}(g_{1})+L^{\rm kin}(g_{2})-tr(g_{1}^{-1}
\pa_z g_{1}\pa_{\zb} g_{2}g_{2}^{-1}+g_{1}^{-1}\pa_{\zb}g_{1}
\pa_z g_{2} g_{2}^{-1})\nn
&&\chi (g_{1}g_{2})=\chi (g_{1})+\chi (g_{2})-d\,tr(g_{1}^{-1}dg_{1}\,dg_{2}
g_{2}^{-1})\ \ .
\label{PWid}
\eea 
The boundary value of $g\in G$ 
is \cite{KG}\cite{ES}\cite{KRWZ} 
\beq
 g(\tau)\ =\ c_G\,c_H\ =\ n\uf n^{-1\,}p\ul p^{-1}(\tau)
\label{gb}
\eeq
where, again, $\uf$ and $\ul$ have no $\tau$-dependence. We 
can write $\uf=e^{2\pi i\la_{G}/k}$ and $\ul=e^{2\pi i\la_{H}/k_H}$, where
$\la_{G}$ and $\la_{H}$ are elements of the Cartan subalgebras 
of the Lie algebras of $G$ and $H$. The single-valuedness 
of path integrals involving the action (\ref{gh}) 
leads to the quantization 
conditions~\cite{KG}\cite{ES}
\beq
\a_{G}(\lambda_{G}) \in  \Z
\label{q1}
\eeq
\beq
\a_{H}(\lambda_{H}) \in  \Z
\label{q2}
\eeq
for any coroots  $\a_{G}$ and $\a_{H}$ of the Lie algebras of $G$ and $H$.
When $H$ is an 
abelian subgroup, the second condition  (\ref{q2}) 
does not apply.

To construct the left-right asymmetrically gauged WZW action for
open strings, we first recall some results for closed strings 
\cite{BS} (see also \cite{Wit}\cite{SfTs}, e.g.). 
Introduce
the gauge fields $A$ and $\tilde{A}$ as 
\bea
A_z=T_a A_z^a, \ \ \ A_{\bar{z}}=T_a A_{\bar{z}}^a,\ \ \ 
\tilde{A}_z=\tilde{T}_a A_z^a,
\ \ \ \tilde{A}_{\bar{z}}=\tilde{T}_a A_{\bar{z}}^a\ \ .
\label{A1}
\eea
Notice that $A_z$, $A_{\bar{z}}$ are independent gauge
fields. On the other hand, 
$\tilde{A}_z$, $\tilde{A}_{\bar{z}}$ are related to
$A_z$, $A_{\bar{z}}$, respectively -- they only differ because 
$T_a$, $\tilde{T}_a$ indicate possibly different embeddings of
a generator of the Lie algebra of the subgroup $H$. 
These must obey the constraints \cite{BS}
\bea
tr(T_a[T_b,T_c])=tr(\tilde{T}_a[\tilde{T}_b,\tilde{T}_c]),\ \ \ \ 
tr(T_a T_b)=tr(\tilde{T}_a \tilde{T}_b)\ \ .
\label{cons}
\eea
Actually, the second constraint guarantees the first, and means the 
two embeddings of $H$ in $G$ must have the same index. 

The left-right asymmetrical transformations are defined as \cite{BS}
\beq
g\rightarrow mg{\tilde{m}}^{-1}
\label{trans}
\eeq
\bea
A_z\rightarrow m{A_z}m^{-1}+\pa_{z}m\,m^{-1},\nn
A_{\zb}\rightarrow mA_{\zb}m^{-1}+\pa_{\zb}m\,m^{-1}, \nn
{\tilde{A}}_z\rightarrow
\tilde{m}{\tilde{A}}_z\tilde{m}^{-1}+\pa_{z}\tilde{m}\,\tilde{m}^{-1},\nn
{\tilde{A}}_{\zb}\rightarrow
\tilde{m}{\tilde{A}}_{\zb}\tilde{m}^{-1}+\pa_{\zb}\tilde{m}\,\tilde{m}^{-1}.
\label{trans1}
\eea 
We write $m\in H_L\subset G$, and ${\tilde{m}}\in H_R\subset G$, for obvious 
reasons. 

The action \cite{BS}
\bea
S^{G/H}=S
+\frac{k}{4\pi}\int_\Sigma d^2z\,tr\left\{
2A_{z}\pa_{\zb}gg^{-1}-2\tilde{A}_{\zb}g^{-1}\pa_{z}g+
2A_z g\tilde{A}_{\zb} g^{-1}-A_z A_{\zb}-\tilde A_z \tilde A_{\zb}\ .
\right\}
\label{asym}
\eea 
is invariant under the left-right asymmetrical transformations (\ref{trans}) 
and (\ref{trans1}). 
In \cite{AT}, this asymmetrically gauged WZW action 
(\ref{asym}) for closed strings was used to remove 
axial $U(1)$ subgroup degrees of freedom. 

We introduce the subgroup valued world sheet fields 
$U,U^{'}\in H_L\subset G$ and $\tilde{U},\tilde{U}^{'}\in H_R\subset G$ as
\bea
A_z=\pa_z UU^{-1},\ \ \ \ A_{\bar{z}}=\pa_{\bar{z}}U^{'}{U^{'}}^{-1}\nn
\tilde{A}_z=\pa_z \ut\ut^{-1},\ \ \ \ \tilde{A}_{\bar{z}}
=\pa_{\bar{z}}{\ut}^{'}{\ut}^{'-1}\ .
\label{AAt}
\eea
Here $U$ and $U^{'}$ are independent subgroup valued fields. 
They are, however, 
related to $\tilde{U}$ and $\tilde{U}^{'}$, respectively, by relations 
similar to (\ref{A1}). That is, the generators of the algebras 
of the subgroups containing $U,U^{'}$  are $\{T_a\}$, while 
those for $\tilde U,\tilde{U^{'}}$  are $\{\tilde T_a\}$. 
The action (\ref{asym}) can be written as
\beq
S^{G/H}=S(U^{-1}g\tilde{U}^{'})-S(U^{-1}U^{'})
\label{apart}
\eeq
which is gauge invariant, since 
(\ref{trans1}) is realized by $U\rightarrow mU$,  $U^{'}\rightarrow mU^{'}$,
$\tilde U\rightarrow \tilde m\tilde U$
and 
$\tilde U^{'}\rightarrow \tilde m\tilde U^{'}$. 

If $H= U(1)$, two special cases occur when the generators 
$\tilde{T}_a=T_a$ and 
$\tilde{T}_a=-T_a$.\,
Then (\ref{asym}) is the action 
of the vectorially and axially gauged WZW model, respectively. 

Let us now turn to the open string case. By exploiting (\ref{Sopen})
for the vectorially gauged WZW model for the open string, (\ref{apart})
shows that the action for the asymmetrically gauged WZW action 
can be constructed as
\bea
 S^{G/H}&=&\frac{k}{4\pi}\left(\int_\Sigma d^2z\,L^{\rm kin}
({U^{-1}g\tilde{U}^{'}})
  +\int_{B}\chi({U^{-1}g\tilde{U}^{'}})-\int_D \om ({U^{-1}g\tilde{U}^{'}})
\right)\nn
 &-&\frac{k_H}{4\pi}\left(\int_\Sigma d^2z\,L^{\rm kin}(U^{-1}U^{'})
  +\int_{B}\chi(U^{-1}U^{'})-\int_D \om (U^{-1}U^{'})\right)\ , 
\label{oa1}
\eea
where on the boundary and the auxiliary disk $D$, 
the fields ${U^{-1}g\tilde{U}^{'}}$, $U^{-1}U^{'}$
take values on the conjugacy 
classes of $G$ and $H$
\bea
{U^{-1}g\tilde{U}^{'}}&=&
(s^{-1}{c_H}^{-1}n)\uf (s^{-1}{c_H}^{-1}n)^{-1}\nn
U^{-1}U^{'}&=& (s^{-1}p){\ul}^{-1} (s^{-1}p)^{-1}
\label{uu'}
\eea
Here we choose this parametrization so that the boundary value of $g$ 
takes the following form
\beq
g(\tau)\ =\ c_G\,c_H\,b_s =\ n\uf n^{-1}\,p\ul p^{-1}s{\tilde{s}}^{-1}(\tau)
\label{abc}
\eeq
where $\uf,n\in G$, $\ul,p\in H$,\ $s \in H_L$, $\tilde{s} \in H_R$
and $b_s = s{\tilde{s}}^{-1}$. 
The fixed group elements $\uf$, $\ul$ parametrize the conjugacy classes
$\{ n\uf n^{-1}\,|\, n\in G \}$, $\{ p\ul p^{-1}\,|\, p\in H\}$. 
$s \in H_L$ and $\tilde s\in H_R$ represent the boundary values of the fields 
$U^{'}$ and ${\tilde U}^{'}$.
This boundary condition (\ref{abc}) allows the  
symmetry (\ref{trans}) to be preserved on the boundary. That is, 
$n\uf n^{-1}p\ul p^{-1}s\tilde{s}^{-1}(\tau)\rightarrow 
(mn)\uf (mn)^{-1}(mp)\ul (mp)^{-1}(ms)(\tilde m\tilde{s})^{-1}$. 

We also mention that the boundary conditions 
for the vectorially gauged model (\ref{gb}) can be 
recovered easily from (\ref{abc}), by putting $\tilde s = s$.

Under the parametrization (\ref{uu'}), the Alekseev-Schomerus two-forms
$\om ({U^{-1}g\tilde{U}^{'}})$ and $\om (U^{-1}U^{'})$,   
defined on the conjugacy classes of $G$ and $H$, are \cite{AS}
\bea
\om ({U^{-1}g\tilde{U}^{'}})&=&
tr\left\{(s^{-1}{c_H}^{-1}n)^{-1}d(s^{-1}{c_H}^{-1}n)
\uf (s^{-1}{c_H}^{-1}n)^{-1}d(s^{-1}{c_H}^{-1}n)\uf^{-1}
 \right\}\nn
\om (U^{-1}U^{'})&=&tr
\left\{(s^{-1}p)^{-1}d(s^{-1}p)\ul^{-1}(s^{-1}p)^{-1}d(s^{-1}p)\ul
  \right\}
\label{ww}
\eea

On the boundary,
the transformation (\ref{trans}) is reduced
to $n\rightarrow mn$, $p\rightarrow mp$,$s\rightarrow ms$, 
and $\tilde s\rightarrow 
\tilde m\tilde s$, so
$\om ({U^{-1}g\tilde{U}^{'}}), \om (U^{-1}U^{'})$ 
are gauge invariant under the transformations
(\ref{trans}) and (\ref{trans1}). As a result,
the action  (\ref{oa1}) is manifestly 
invariant under continuous deformations of the embedding
of the auxiliary disk inside the conjugacy class.

The action  (\ref{oa1}) is still sensitive to a topological change
in the embedding of the auxiliary disk. To ensure such a change
has no observable effect, the induced change in the action
should be an integer multiple of $2\pi$. This constraint leads
to the quantization conditions (\ref{q1}) and (\ref{q2}).
However, under the topological change
in the embedding of the auxiliary disk,
the factor $s{\tilde s}^{-1}$ does not lead to any quantization condition; 
this can be seen from the construction of the action (\ref{oa1})
and the parametrization for (\ref{uu'}).

By exploiting (\ref{PWid}) and (\ref{ww}), 
the left-right asymmetrically gauged WZW action for
open strings (\ref{oa1}) can be written as
\bea
S^{G/H}&=&\frac{k}{4\pi}\left(\int_\Sigma d^2z\,L^{\rm kin}(g)
+\int_B\chi(g)\right)\nn
&+&\frac{k}{4\pi}\int_\Sigma d^2z\,tr\left\{
2A_{z}\pa_{\zb}gg^{-1}-2\tilde{A}_{\zb}g^{-1}\pa_{z}g+
2A_z g\tilde{A}_{\zb} g^{-1}-A_z A_{\zb}- \tilde A_z \tilde A_{\zb}
\right\}\nn
&-&\frac{k}{4\pi}\int_D\,\Big\{
\Omega(n;\uf)+\Omega(p;\ul)
+tr\Big(s^{-1}ds\tilde{s}^{-1}d\tilde{s}\nn
&&
-{c_G}^{-1}dc_G(dc_H{c_H}^{-1}+{c_H}db_s{b_s}^{-1}{c_H}^{-1})
-{c_H}^{-1}dc_H db_s{b_s}^{-1}
\Big)\Big\}
\label{agh}
\eea
The two-form in the last bracket has the pullback 
$g^*\chi$ $=(n\uf n^{-1}p\ul p^{-1}s\tilde{s}^{-1})^* \chi$
as its exterior derivative.
When $\tilde s = s$, the action (\ref{agh}) is reduced
to the vectorially gauged WZW action for open string 
(\ref{gh}).
Thus the asymmetrically gauged WZW action for open strings 
can be determined by the action (\ref{agh}), with boundary condition
(\ref{abc}),  plus the quantization conditions  
(\ref{q1}) and (\ref{q2}).

Now, consider the axially gauged WZW model, with the axial $U(1)$ subgroup 
chosen by taking $\tilde{T}_a=-T_a$. We have
\bea
S^{G/U(1)_{A}}&=&\frac{k}{4\pi}\left(\int_\Sigma d^2z\,L^{\rm kin}(g)
+\int_B\chi(g)\right)\nn
&+&\frac{k}{2\pi}\int_\Sigma d^2z\,tr\left\{
A_{z}\pa_{\zb}gg^{-1}+A_{\zb}g^{-1}\pa_{z}g-
A_z g A_{\zb} g^{-1}-A_z A_{\zb}
\right\}\nn
&-&\frac{k}{4\pi}\int_D \Big\{
\Omega(n;\uf)
-tr\Big((n\uf n^{-1})^{-1} 
d(n\uf n^{-1})d(s\tilde{s}^{-1})(s\tilde{s}^{-1})^{-1}
\Big)\Big\}\ .
\label{au1}
\eea
Here the subscript $A$ stands for axial. 
The boundary value for the field $g$ is reduced to
$g(\tau)=n\uf n^{-1}s\tilde{s}^{-1}\ul$. For the case of $H=U(1)$,
the second quantization condition (\ref{q2}) does not apply, so 
$\ul$ does not get quantized.
We can rescale $g\rightarrow g\ul^{-1}$, then the boundary value for new $g$
is $g(\tau)=n\uf n^{-1}s\tilde{s}^{-1}(\tau)$\footnote{
Another justification for this boundary condition 
arises later 
in the parafermion theory, where  B-type D-branes are described by 
the boundary states $|Bj\rangle$ labelled by a single quantum number 
$j$ \cite{MMS}. We show below that B-type D-branes are related to 
the D-branes in the axially gauged WZW model. Accordingly, only one 
fixed and quantized group element $\uf$ is left in (\ref{abc}) to 
characterize the boundary condition.}.
Under this rescaling
the action (\ref{au1}) keeps invariant. 

In order to transform to the  T-dual theory, 
we follow the argument for the presence of the Killing
symmetries associated to the Cartan subalgebra of
the subgroup \cite{EK}, \cite{EK1}, and parametrize $g=he^{iXT^{0}}$
with $0\leq X\leq 2\pi$.  
The action (\ref{au1}) becomes
\bea
S^{G/U(1)_{A}}&=&\frac{k}{4\pi}\left(\int_\Sigma d^2z\,L^{\rm kin}(h)
+\int_B\chi(h)-\int_D d^2z\,\Omega(n;\uf)\right)\nn
&+&\frac{k}{4\pi}\int_\Sigma d^2z\,tr\Big\{\pa_{z}X\,\pa_{\zb}X+
2i\pa_{z}X\, A_{\zb}-2i\pa_{\zb}X (U^{0}_z - A_z M)\nn
&&
-2[(1+M)A_z A_{\zb} -A_z V^{0}_{\zb}-A_{\zb}U^{0}_z]
\Big\}\ ,
\label{au2}
\eea
and\ the\ boundary\ conditions\ are\ $h(\tau)=n\uf n^{-1}(\tau)$,\ 
$e^{iX(\tau)T^0}=s\tilde{s}^{-1}(\tau)$, and
$U^{0}_z = tr(T^0 h^{-1}\pa_{z}h)$, 
$V^{0}_{\zb}= tr(T^0 \pa_{\zb}hh^{-1})$, $M = tr(T^0 h T^0 h^{-1})$. 
To start the dualization procedure we
rewrite the partition function of the action (\ref{au2}) as 
\cite{ABB}-\cite{BL}
\beq
Z(h,A)=\int\,DyD\tilde{X}e^{i\tilde{S}(h,A,y,\tilde{X})}
\eeq
with 
\bea
\tilde{S}(h,A,y,\tilde{X})&=&\frac{k}{4\pi}\Big[\int_\Sigma d^2z\,
L^{\rm kin}(h)
+\int_B\chi(h)-\int_D \Omega(n;\uf)\Big]\nn
&+&\frac{k}{4\pi}\int_\Sigma d^2z\,\Big\{y_{z}y_{\zb}+2iy_{z}A_{\zb}
-2iy_{\zb}(U^{0}_z -A_z M)+y_{z}\pa_{\zb}\tilde{X}-y_{\zb}\pa_{z}\tilde{X}\nn
&&
-2\Big[(1+M)A_z A_{\zb} -A_z V^{0}_{\zb}-A_{\zb}U^{0}_z\Big]\Big\}\nn
&&
+\frac{k}{4\pi}\int_{\pa{\Sigma}}\tilde{X}y(\tau)\,d\tau
\eea
The T-dual action can be obtained by performing the $y$ functional
integration. The presence of the boundary induces
a crucial subtlety \cite{DO}. Since the $y$ integral is
ultralocal we can use \cite{DO}
\beq
\int_\Sigma Dy\, e^{i\tilde{S}}=\int_\Sigma Dy \exp\Big(i\tilde{S}-
\frac{ik}{4\pi}
\int_{\pa{\Sigma}} \tilde{X}y(\tau)d\tau\Big)\cdot 
\int_{\pa{\Sigma}}Dy\exp\Big(
\frac{ik}{4\pi}
\int_{\pa{\Sigma}} \tilde{X}y(\tau)d\tau\Big)
\eeq
Then the $y$ integral on the boundary $\pa{\Sigma}$ results
in a $\d$ function which imposes the boundary condition for  
$\tilde{X}$
\beq
\tilde{X}(\tau)=0
\label{xbc}
\eeq
The boundary condition for the original $U(1)$ coordinate $X$
is $e^{iX(\tau)T^0}=s\tilde{s}^{-1}(\tau)$, 
but for the T-dual coordinate $\tilde{X}$,
it becomes $\tilde{X}(\tau)=0$. In the open string case, axial-vector
duality is realized with these different
boundary conditions.

After integrating out the bulk $y_{z}$, $y_{\zb}$ fields, we get
\beq
Z(h,A)=\int\,D\tilde{X}\exp\Big(i\tilde{S}(h,A,\tilde{X})\Big)
\eeq
with
\bea
\tilde{S}(h,A,\tilde{X})&=&\frac{k}{4\pi}\Big[\int_\Sigma d^2z\,
L^{\rm kin}(h)
+\int_B\chi(h)-\int_D \Omega(n;\uf)\Big]\nn
&+&\frac{k}{4\pi}\int_\Sigma d^2z\,\Big\{\pa_{z}\tilde{X}\,\pa_{\zb}\tilde{X}
-2i\pa_{z}\tilde{X}\,A_{\zb}
-2i\pa_{\zb}\tilde{X}(U^{0}_z -A_z M)\nn
&&
+2\Big(A_z V^{0}_{\zb}-A_{\zb}U^{0}_z + A_z A_{\zb} M -  A_z A_{\zb}
\Big)\Big\}\ .
\eea
If we define the T-dual field $\tilde{g}=h e^{i\tilde{X}T^0}$,
we have 
\bea
\tilde{S}&=&\frac{k}{4\pi}\left(\int_\Sigma d^2z\,L^{\rm kin}(\tilde g)
+\int_B\chi(\tilde g)\right)\nn
&+&\frac{k}{2\pi}\int_\Sigma d^2z\,tr\left\{
A_{z}\pa_{\zb}\tilde{g}\tilde{g}^{-1}-A_{\zb}\tilde{g}^{-1}\pa_{z}\tilde{g}
+A_z \tilde{g}A_{\zb} \tilde{g}^{-1}-A_z A_{\zb}
\right\}\nn
&-&\frac{k}{4\pi}\int_D \Omega(n;\uf)\ .
\label{vu}
\eea
The boundary condition for the T-dual group-valued field $\tilde{g}$ is
\beq
 \tilde{g}(\tau)=n\uf n^{-1}(\tau)\ .
\eeq
Thus the gauged WZW action with vector $U(1)$ gauge subgroup
(\ref{gh}) is recovered, with the choice $\ul=1$.

Here we should point out that the angle $X$ was originally
normalized to take values in $[0,2\pi]$. The duality 
transformation changes this range: the
dual angle $\tilde{X}$ takes values in $[0,\frac{2\pi}{k}]$.
Thus we see that the axially gauged WZW model is T-dual to the 
$\Z_{k}$ orbifold of the vectorially gauged WZW model.

\section{A-branes in the $SU(2)/U(1)$ coset model}

In the $SU(2)/U(1)$ coset model, the boundary condition 
of the group-valued field $g$ (\ref{gb}) is 
reduced to 
\beq
 g(\tau)=n\uf n^{-1}(\tau)\ul
\label{gb1}
\eeq
where $\ul$ is an arbitrary but fixed group element of the $U(1)$
subgroup that is not quantized. The gauged WZW action for open
strings obtained by gauging away a vector $U(1)$
subgroup (\ref{gh}) is 
\bea
S^{SU(2)/U(1)}&=&S^{SU(2)}-\frac{k}{4\pi}\int_D \Omega(n;\uf)\nn
&+&\frac{k}{2\pi}\int_\Sigma d^2z\,tr\left\{
A_{z}\pa_{\zb}gg^{-1}-A_{\zb}g^{-1}\pa_{z}g+A_z gA_{\zb} g^{-1}-A_z A_{\zb}
\right\}
\label{action}
\eea
We use the following 
two parametrizations of $SU(2)$ group elements 
\beq
 g = e^{i(\tilde{\phi}+{\phi})\frac{\sigma_{3}}{2}}
  e^{i\theta\sigma_{1}}
e^{i(\tilde{\phi}-{\phi})\frac{\sigma_{3}}{2}} = e^{i\psi 
\vec{n}\cdot\vec{\si}}\ \ ,
\label{para}
\eeq 
following \cite{MMS}. 
Here $0\leq\t\leq \frac{\pi}{2}$, $0\leq\phi,\tilde{\phi}\leq 2\pi$,
$0\leq\psi\leq \pi$. We choose the spherical coordinates as
\beq 
\vec{n}\cdot\vec{\si}=(\sin\eta\cos\varphi\si_{1},\, 
\sin\eta\sin\varphi\si_{2},\,
\cos\eta\si_{3})
\label{para1}
\eeq
where $0\leq\eta\leq \pi$, $0\leq\varphi\leq 2\pi$. The relation
between the two parametrizations is
\beq 
\cos\psi=\cos\t\cos\tilde{\phi}
\label{rela1}
\eeq
and
\beq 
\varphi=2\pi-\phi,\ \ \ \
\tan\eta=\tan\t/\sin\tilde{\phi}
\label{rela2}\ .
\eeq
A D-brane in the $SU(2)$ group manifold is an 
$S^2$ described in spherical coordinates by  \cite{AS}
\beq 
\psi_{j}=\frac{2j}{k}\pi\ .
\label{q3}
\eeq
$\psi_{j}$ characterizes different spherical D2-branes
with $j\in\frac{1}{2}\Z$, $0\leq j\leq \frac{k}{2}$, and  
eq.(\ref{q3}) can be derived from the quantization condition (\ref{q1}).
Under the transformation $g\rightarrow g\ul^{-1}$, the vectorially gauged
WZW action for open strings (\ref{action}) is invariant, but
the boundary condition turns to be $g(\tau)=n\uf n^{-1}(\tau)$.
Inserting (\ref{para}) into (\ref{2form}), we have \cite{SF}
\beq 
\Omega(n;\uf)=B+F=2\cos^{2}\t\, d\phi\wedge d\tilde{\phi}
\label{b+f}
\eeq
where $F$ is the gauge field strength on D-branes 
(not the field strength of the auxiliary gauge fields $A_z$ and $A_{\zb}$
in (\ref{action}), whose role is to gauge away a $U(1)$
subgroup). When we parametrize the gauge fields as 
\beq 
A_z=-\frac{i}{2}\si_{3}A,\ \ \ \ A_{\zb}=-\frac{i}{2}\si_{3}\bar{A}
\label{gauge}
\eeq
and insert (\ref{para}) and (\ref{gauge}) into (\ref{action}), we find 
$$
S^{SU(2)/U(1)_{V}}=k\int d^2z 
\left(\pa_{z}\t\, \pa_{\zb}\t+\frac{1}{\tan^2\t}\,
\pa_{z}\tilde{\phi}\,\pa_{\zb}\tilde{\phi}\right)\qquad\qquad\qquad\qquad
\qquad
$$
\beq
+ k\int d^2z\sin^{2}\t\left(A+\pa_{z}\phi+\frac{1}{\tan^2\t}
\pa_{z}\tilde{\phi}\right)\left(\bar{A}+\pa_{\zb}\phi-\frac{1}{\tan^2\t}
\pa_{\zb}\tilde{\phi}\right)
\label{va1}
\eeq
where the subscript $V$ indicates that we have gauged the
vector $U(1)$ subgroup away. In deriving (\ref{va1}) 
we have omitted an irrelevant overall factor,
and used the special form for the NS 2-form $B$ field.
In the $SU(2)$ group manifold, the NS field strength 3-form is
\beq 
\chi(g)=\frac{1}{3}tr(dgg^{-1})^3=-2 d\phi\wedge d\tilde{\phi}\wedge 
d(\sin^{2}\t)
\eeq
and we can choose locally either  $B=2\cos^{2}\t d\phi\wedge d\tilde{\phi}$
or $B=-2\sin^{2}\t d\phi\wedge d\tilde{\phi}$ due to the
coordinate singularity at $\t = \pi/2$ and $\t = 0$ points.\footnote{\,In
spherical coordinates, we can choose either 
$B=k(\psi-\frac{1}{2}\sin2\psi)\sin\eta d\eta\wedge d\varphi$ 
or $B=k(\psi-\pi-\frac{1}{2}\sin2\psi)\sin\eta d\eta\wedge d\varphi$.
In the context of 10D supergravity, when we consider D3-brane probes 
in the background of $k$ coincident NS5-branes, the different
choices of $B$ result in different numbers of D1-branes. This 
is nothing other than the brane creation effect \cite{JGZ}.}  
When gauging the vector $U(1)$ subgroup, we 
choose $B=2\cos^{2}\t d\phi\wedge d\tilde{\phi}$. From 
(\ref{b+f}) we then see that the gauge field strength $F$ living at the
end points of open strings is zero.

Integrating $A$ and $\bar{A}$ out removes the last term
in (\ref{va1})  completely, 
and produces a dilaton proportional to $\log\sin\t$. 
The resulting $\si$-model action for open strings is
\beq
S^{SU(2)/U(1)_{V}}=k\int d^2z\, \frac{1}{1-\rho'^{2}}(\pa_{z}{\rho'}\,
\pa_{\zb}{\rho'}+\rho'^{2}\pa_{z}\tilde{\phi}\,\pa_{\zb}\tilde{\phi})
\label{va2}
\eeq
\beq
g_{s}(\rho')=e^{\Phi}=g_{s}(0)(1-\rho'^{2})^{-\frac{1}{2}}
\label{vg}
\eeq
where we have chosen  $\rho=\sin\t$ and $\rho'=\sqrt{1-\rho^{2}}$. 
The topology of the resulting target space is a 
disk.\footnote{\,In order to see the relation between our geometrical
interpretation of the D-branes and that in \cite{MMS}, we adopt
the same notation as in  \cite{MMS}.} Recall that the NS 2-form $B$ field in 
(\ref{va1}) is $B=2\cos^{2}\t\, d\phi\wedge d\tilde{\phi}$. This tells us 
there is a singular point at $\t =0$, i.e., $\rho'=1$. In 
(\ref{va2}) there is a curvature singularity at $\rho'=1$, so this
curvature singularity results from the particular choice of 
NS 2-form $B$ field.

In the action (\ref{va2}) there is no gauge field at the end
points of the open string, and its geometry is conformally
flat. Therefore, the boundary condition surviving from
those of D-branes in the $SU(2)$ parent theory should remain intact. That is,
it is compatible with the boundary condition obtained by
variation of the action (\ref{va2}). Combining
(\ref{rela1}) and (\ref{q3}), the surviving boundary condition
(removing the $\phi$ coordinate) is 
\beq 
\cos\psi_{j}=\cos\t\cos\tilde{\phi}\ \ .
\label{bc1}
\eeq
This characterizes D-branes in the gauged WZW model with a 
vector $U(1)$ gauge subgroup. If we define $\rho'_{j}
=\sqrt{1-\rho^{2}_{j}}=\cos\psi_{j}$, the boundary condition
(\ref{bc1}) can be recast into the form 
\beq 
\cos\tilde{\phi}=\frac{\rho'_{j}}{\rho'}
\label{bc2}
\eeq
or 
\beq
\tilde{\phi}-\arccos\sqrt{\frac{1-\rho^{2}_{j}}{1-\rho^{2}}}=0\ .
\label{bc3}
\eeq
(\ref{bc2})
shows that that for fixed $j$, the shape of an A-brane is a 
straight line in the disk $(\rho',\tilde{\phi})$,   
as depicted in Figure 1.

%%%%%%% Fig. 1 %%%%%%%%%%%%%%%%
\begin{figure}[htbp]
\begin{center}
\includegraphics*[scale=.60]{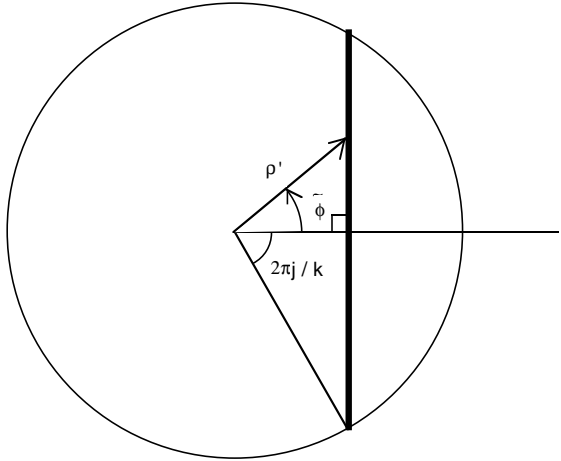}
\vspace{-0.7cm}
\end{center}
\caption{The shape of a D1-brane for fixed $j$ described by (\ref{bc2}).}
\end{figure}
%%%%%%%%%%%%%%%%%%%%%%%%%%%%%%%

The boundary condition for open strings (\ref{bc1}) obtained
from the vectorially gauged WZW model can be verified from the scattering
amplitude between the boundary states and closed string
states. The action (\ref{va2}) indicates that the wave functions 
on the disk $(\rho',\tilde{\phi})$ are  $SU(2)$ wave functions that are
invariant under translations of $\phi$. So, the wave functions
of the parafermion theory are\,\footnote{\, 
The authors of \cite{MMS} argued that the wave functions
of the parafermion theory are invariant under translations of 
$\tilde{\phi}$. We differ. It seems that in \cite{MMS} 
the axially gauged WZW model was mistakenly used 
to describe the $SU(2)/U(1)$ coset model.} 
\bea
\Psi_{j,m}(\t,\tilde{\phi})&\ =\ &\langle j,m|g|j,m\rangle\ 
=\ e^{ i2m\tilde{\phi}}
\langle j,m|e^{i2\t J^1}|j,m\rangle\nn
&\ =\ &\frac{1}{\sqrt{2j+1}}\langle \t,\tilde{\phi}|g|j,m,m\rangle_{PF}
\eea
where the $|j,m\rangle$ are a basis for the spin $j$ representation of 
$SU(2)$.

The Cardy boundary states for A-branes are \cite{MMS}
\beq 
|A,\hat{j},\hat{n}\rangle_{C}\ =\ \sum_{(j,n)\in PF(k)}
\frac{S_{\hat{j}\hat{n}}^{PF jn}}
{\sqrt{S_{00}^{PF jn}}}|A,j,n\rangle\rangle\ .
\eeq
The scattering amplitude between A-brane boundary states and closed string
states is \cite{FFFS}\cite{MMS} 
\bea
_C{\langle}A,\hat{j},\hat{m}|\t,\tilde{\phi}\rangle&\sim &\sum_{j}S_{\hat{j},j}
e^{\frac{i4\pi m\hat{m}}{k}}\Psi_{j,m}(\t,\tilde{\phi})^{*}
=\sum_{j}\sin\Big[\frac{2\hat{j}\pi}{k+2}(2j+1)\Big]\nn
&&
\cdot
\Big[\sum_{m=-j}^{j}\langle j,m|e^{i2(\tilde{\phi}-\frac{2\pi\hat{m}}{k})J^3}
e^{i2\t J^1}|j,m\rangle\Big]^*
\label{ampli}
\eea
where we have assumed $k\gg 1$. 

To calculate  $_C{\langle}A,\hat{j},\hat{m}=0|\t,\tilde{\phi}\rangle$, we
exploit the relation
\beq 
\sum_{m=-j}^{j}\langle j,m|e^{i2\tilde{\phi}J^3}
e^{i2\t J^1}|j,m\rangle\ =\ \frac{\sin(2j+1)\tilde{\psi}}{\sin\tilde{\psi}}
\label{jm}
\eeq
with 
\beq 
\cos\tilde{\psi}=\cos\t\cos\tilde{\phi}\ .
\label{good}
\eeq
Inserting (\ref{jm}) and (\ref{good}) into  (\ref{ampli}), we get
\beq 
_C{\langle}A,\hat{j},\hat{m}=0|\t,\tilde{\phi}\rangle\ \sim\  
\d (\tilde{\psi}-
\frac{2\hat{j}\pi}{k})
\label{m=0}
\eeq
Eqs. (\ref{good}) and (\ref{m=0})  show that the shape of an A-brane
in the parafermion theory is described by
\beq 
\cos\frac{2\hat{j}\pi}{k}=\cos\t\cos\tilde{\phi}\ \ .
\eeq
This agrees with  (\ref{bc1}),  obtained above from the vectorially gauged
WZW model for open strings.

Since $0\leq j \leq \frac{k}{2}$, the boundary condition (\ref{bc2})
shows that
the $k-1$ spherical D2-branes
in the $SU(2)$ group manifold turn into  $k-1$ D1-branes. 
The two original D0-branes in $SU(2)$ are projected to D0-branes 
in the disk $(\rho',\tilde{\phi})$.  All this is illustrated
in Figure 2.

%%%%%%% Fig. 2 %%%%%%%%%%%%%%%%
\begin{figure}[htbp]
\begin{center}
\includegraphics*[scale=.60]{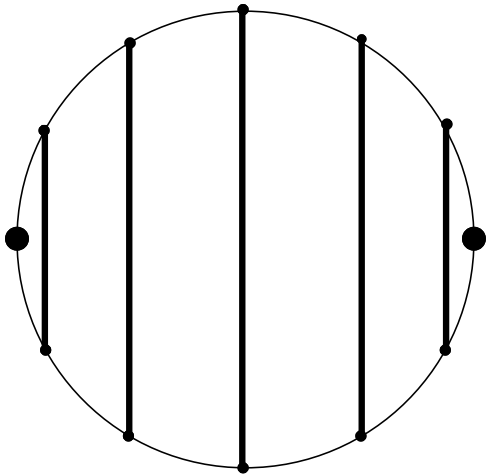}
\vspace{-0.7cm}
\end{center}
\caption{In the level $k(=6)$ case, there are $k-1(=5)$ straight D1-branes
and two D0-branes on the disk.}
\end{figure}
%%%%%%%%%%%%%%%%%%%%%%%%%%%%%%%

The vectorially gauged WZW action for open strings is invariant under
the transformation $g\rightarrow g\ul^{-1}$ where the fixed $U(1)$ 
group element  $\ul$ is not quantized, but the
boundary condition changes. In the parametrization
 (\ref{para}), this $U(1)$ transformation is realized as 
$\phi \rightarrow\phi-\b$, $\tilde{\phi} \rightarrow\tilde{\phi}+\b$.
To match the result in $\si$-model of open strings with that
in CFT, Eq.(\ref{ampli}) shows that the boundary condition should be
modified as 
\beq 
\cos\psi_{j}=\cos\t\cos(\tilde{\phi}-\frac{n\pi}{k})
\label{zkb}
\eeq
where $-k+1\leq n \leq k$, that is, $U(1)$ symmetry is broken
to $\Z_{2k}$ symmetry. From CFT points of view, when 
the pairs subject to a constraint
$2j+n=0$ mod 2, then half of the number of  $n$ is eliminated,
and we are left with $\Z_{k}$ symmetry. Eq.(\ref{zkb}) shows
that when $j=0$ there are $k$ D0-branes at one of the $k$
special points around the disk, and $\left(k\atop 2\right)$ D1-brane
states with $j=\frac{1}{2},1,\ldots, 
\frac{1}{2}[\frac{k}{2}]$ which stretch between two of the special
points around the disk separated by $2j$ segments\footnote{\,Eq.(\ref{zkb}) 
shows that two points of  D1-brane
are separated by $2\times\frac{2j\pi}{k}=2j\times\frac{2\pi}{k}$,
that is, $2j$ segments instead of  $4j$ segments.} as shown in
Figure 3.

%%%%%%% Fig. 3 %%%%%%%%%%%%%%%%
\begin{figure}[htbp]
\begin{center}
\includegraphics*[scale=.60]{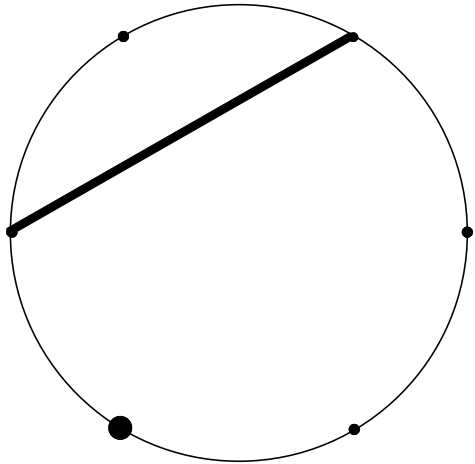}
\vspace{-0.7cm}
\end{center}
\caption{The geometry of A-branes for fixed $j$ in the $k=6$ case. 
The points marked around the boundary disk indicate the possible D0-branes.}
\end{figure}
%%%%%%%%%%%%%%%%%%%%%%%%%%%%%%%

From the surviving boundary condition (\ref{bc2}) and the 
$\Z_{k}$ symmetry, we have recovered the geometry of A-branes uncovered in 
\cite{MMS}. There this picture was conjectured, mainly from
CFT. The geometrical interpretation of A-branes in the $SU(2)/U(1)$
coset model is that the spherical D2-branes on the $SU(2)$
group manifold are projected to the disk $(\rho',\tilde{\phi})$,
and the resulting shape is a straight line.

\section{B-branes and axial-vector duality}

Since there is a $\Z_{k}$ symmetry in the parafermion theory, it
was argued in \cite{MMS} that the level $k$ $SU(2)/U(1)$
coset model is equivalent to its $\Z_{k}$ orbifold.
A-branes in the orbifold theory are constructed from A-branes
in its covering theory by adding the images under $\Z_{k}$ 
so that the configuration is $\Z_{k}$ invariant. Then T-duality
maps these branes to B-branes in the original 
theory.\footnote{\, T-duality was also used to compute B-type branes 
in \cite{FSS}, where it was 
applied to Gepner models.}

In the orbifold, the range of values of the angle $\tilde{\phi}$ is
changed to $[0,\frac{2\pi}{k}]$. 
If we introduce a new angle $\hat{\phi}=k\tilde{\phi}$ with
$\hat{\phi}\rightarrow \hat{\phi}+2\pi$, then (\ref{va2}),(\ref{vg}),
then (\ref{bc3}) can be rewritten as
\beq
S^{SU(2)/U(1)_{V}}=\int d^2z \Big(\frac{k}{1-\rho'^{2}}\pa_{z}{\rho'}\,
\pa_{\zb}{\rho'}+\frac{\rho'^{2}}{k(1-\rho'^{2})}\pa_{z}
\hat{\phi}\,\pa_{\zb}\hat{\phi}\Big)
\label{open}
\eeq
\beq
g_{s}(\rho')=g_{s}(0)(1-\rho'^{2})^{-\frac{1}{2}}
\eeq
\beq
\left(\hat{\phi}-k\arccos\sqrt{\frac{1-\rho^{2}_{j}}{1-\rho^{2}}}
\,\right)_{\pa\Sigma}=0
\label{hat}
\eeq
We now perform a T-duality transformation for open strings \cite{DO} along 
the circle $S^1$ parametrized by $\hat{\phi}$. In the dual theory, strings 
end on the hypersurface defined by setting the coordinate in the direction 
of the Killing vector equal  to the corresponding component of the 
gauge potential \cite{DO}. We find the action 
\beq
S^{dual}=k\int d^2z\,\frac{1}{1-\rho^{2}}\Big(\pa_{z}{\rho}\,
\pa_{\zb}{\rho}+\rho^{2}\pa_{z}\phi\,\pa_{\zb}\phi\Big)
+\int_{\pa\Sigma}a
\label{da}
\eeq
with 
\beq
g_{s}(\rho)=\sqrt{k}g_{s}(0)(1-\rho^{2})^{-\frac{1}{2}}
\eeq
and
\beq
a=-k\arccos\sqrt{\frac{1-\rho^2 _j}{1-\rho^2}}\,d\phi
\label{gau}
\eeq
where $\phi$ is the coordinate dual to $\hat\phi$. 

By the variation of the action (\ref{da}), the boundary
condition is
\bea
\left(\pa_{\si}\rho + \sqrt{1-\rho^2 _j}\frac{\rho}{\sqrt{\rho^2 _j-\rho^2}}
\pa_{\tau}\phi\right)_{\si=0,\pi}&=&0\nn
\left(\pa_{\si}\phi - \sqrt{1-\rho^2 _j}\frac{1}{\rho\sqrt{\rho^2 _j-\rho^2}}
\pa_{\tau}\rho\right)_{\si=0,\pi}&=&0
\label{bcb}
\eea
where the coordinates are related by  
$\pa_z=\pa_\tau-\pa_\sigma,\ \pa_{\zb}=\pa_\tau+\pa_\sigma$.
Eq.(\ref{bcb}) describes a D2-brane disk. The gauge field strength on
this D2-brane is
\beq
F\ =\ da\ =\ 
k\frac{\rho}{1-\rho^2}\sqrt{\frac{1-\rho^2 _j}{\rho^2 _j-\rho^2}}\,
d\rho\wedge d\phi\ .
\label{gfs}
\eeq
This shows that the D2-brane disk should have radius $\rho=\rho_j=\sin
\frac{2j\pi}{k}$ and be centered, as illustrated in Figure 4. 

%%%%%%% Fig. 4 %%%%%%%%%%%%%%%%
\begin{figure}[htbp]
\begin{center}
\includegraphics*[scale=.60]{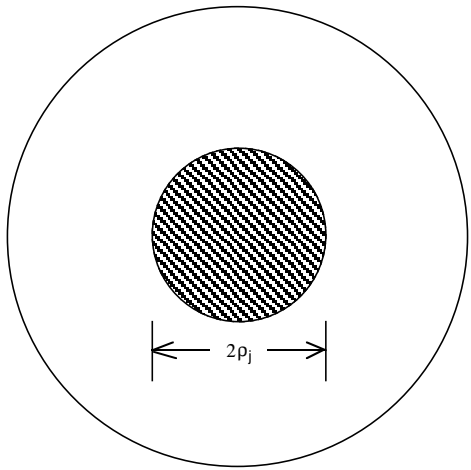}
\vspace{-0.7cm}
\end{center}
\caption{The shape of D2-branes with radius $\rho_j=\sin\frac{2j\pi}{k}$.}
\end{figure}
%%%%%%%%%%%%%%%%%%%%%%%%%%%%%%%

Because of this nonzero gauge field strength, the 
B-branes are stable and cannot decay by shrinking or 
displacing off the center of the disk. 
This was shown in \cite{MMS}, where the B-branes were 
described by an effective theory, and the 
classical equations for a disk D2-brane
with a gauge field strength $F$ were considered. A 
fixed flux of $F$ on the B-brane was 
assumed and its value was calculated by 
minimizing the Dirac-Born-Infeld action.
The $F$ they obtained by their minimization procedure
is exactly the same as that in (\ref{gfs}). However, in the
above approach, such minimization is unnecessary. 
The gauge field strength $F$ on the  D2-brane is obtained more directly -- 
from the gauged WZW model, by first gauging away the vector $U(1)$ subgroup,
then orbifolding and transforming to the T-dual theory.

The mass of the B-branes can be calculated from the  
Dirac-Born-Infeld action
\beq
M_{D2}=\int g_{s}^{-1}(\rho)\sqrt{-det(G_{\a\b} + F_{\a\b})}=
\frac{2\pi\sqrt{k}}{g_{s}(0)}\sin\frac{2j\pi}{k}\ .
\eeq
For large $k$ this mass is proportional to 
\beq
M_{j}=\frac{1}{g_{s}(0)\sqrt{k(k+2)}}\sin\frac{(2j+1)\pi}{k}\ .
\eeq
Thus the mass derived from the $\si$-model for open strings matches
that found from CFT.\footnote{\,By variation of the action (\ref{da}),
we can also obtain other D-brane configurations, but their masses
do not match those derived from  CFT.}

Consider the axially gauged WZW action for open strings (\ref{au1}).  
If we insert (\ref{gauge}) and  (\ref{para}) into (\ref{au1}), we
get
$$ 
S^{G/U(1)_{A}}\ =\ k\int d^2z\,\Big(\pa_{z}\t\,\pa_{\zb}\t+\tan^2\t\,
\pa_{z}\phi\,\pa_{\zb}\phi\qquad\qquad\qquad$$ 
$$
\qquad\qquad\qquad+\cos^{2}\t(A+\pa_{z}\tilde{\phi}+\tan^2\t
\,\pa_{z}\phi) (\bar{A}+\pa_{\zb}\tilde{\phi}-\tan^2\t
\,\pa_{\zb}\phi)\Big)$$
\beq
\qquad\qquad\qquad
-k\oint_{\pa\Sigma}\arccos\sqrt{\frac{1-\rho^2 _j}{1-\rho^2}}\,d\phi\ .
\label{aa}
\eeq
Here we have omitted an irrelevant overall factor, and chosen
the NS B-field as $B=-2\sin^{2}\t d\phi\wedge d\tilde{\phi}$.
Integrating the gauge fields $A$ and $\bar{A}$ out, we obtain
the $\si$-model for B-branes described by (\ref{da})-(\ref{gau}). 
This shows that the B-branes can be realized
in the gauged WZW action with axial $U(1)$ gauge subgroup.

In \cite{MMS},\ the operator $\exp(i\pi\tilde{J}_{0}^1)$ 
was exploited
to define B-type Ishibashi states in the parafermion theory:
\beq
(1\otimes e^{i\pi\tilde{J}_{0}^1})|j\rangle\rangle^{SU(2)}=
\sum_{r=0}^{2k-1}|Bj,r,-r\rangle\rangle^{PF}
\otimes|B,r,-r\rangle\rangle^{U(1)_{k}}
\eeq
The A-type Ishibashi states in the $SU(2)$ group manifold satisfy
the boundary condition
\beq
(J_{n}+R\tilde{J}_{-n})|j\rangle\rangle^{SU(2)}\ =\ 0
\eeq
with $RJ^a=J^a$, for all $a$. The Cardy boundary states constructed
from these maximally symmetric Ishibashi states therefore describe
spherical D2-branes. Since
\beq
e^{i\pi\tilde{J}_{0}^1}\tilde{J}_{n}^{2} 
e^{-i\pi\tilde{J}_{0}^1}=-\tilde{J}_{n}^{2},\ \ 
e^{i\pi\tilde{J}_{0}^1}\tilde{J}_{n}^{3} e^{-i\pi\tilde{J}_{0}^1}
=-\tilde{J}_{n}^{3}
\eeq
we have 
\beq
(J_{n}+R'\tilde{J}_{-n})(1\otimes 
e^{i\pi\tilde{J}_{0}^1})|j\rangle\rangle^{SU(2)}=0
\eeq
with $R'J^a=(-1)^{\delta_{a,1}}J^a$. Then the boundary states  
$(1\otimes e^{i\pi\tilde{J}_{0}^1})|j\rangle\rangle^{SU(2)}$ are also
maximally symmetric. The corresponding Cardy boundary
states describe another set of spherical  D2-branes 
obtainable by a global rotation of the first set. 
Thus the geometrical interpretation of B-branes is that the spherical
D2-branes in the $SU(2)$ group manifold are projected to the 
$(\rho,\phi)$ disk, and their shape is a centered disk with the
radius $\rho=\rho_j=\sin\frac{2j\pi}{k}$. The quantized gauge
field strengths in the original spherical D2-branes of the $SU(2)$ 
parent theory are inherited by these disk B-branes, making 
them stable.

\section{Summary and Discussion}

We have constructed  D-branes in a left-right asymmetrically gauged
WZW model for open strings.  
The asymmetry was restricted to the type induced from 
different embeddings of the gauge
subgroup acting on the left and right of the  group-valued field $g$. 
Boundary
conditions were written for $g$ that preserve this 
symmetry. For a subgroup $H\cong U(1)$, 
the  axially gauged
WZW action was found, and the vectorially gauged WZW action 
was then reproduced by 
implementing
T-duality. We find that 
the boundary condition for the T-dual coordinate 
is $\tilde{X}(\tau)=0$, while 
for the original $U(1)$ coordinate $X$ it 
is $e^{iX(\tau)T^0}=s\tilde{s}^{-1}(\tau)$. 
The axial-vector
duality in the open string case is realized with such a change of the 
boundary condition.
We conclude that the axially gauged WZW model
is T-dual to a $\Z_k$ orbifold of the vectorially gauged WZW model. 
This is an example of an axial-vector duality in open strings.

From the $SU(2)/U(1)$ coset model, described 
by the vectorially gauged WZW model, 
we found a vanishing
gauge field on D-branes. Since the geometry of the 
disk is conformally flat and the gauge field on D-branes
vanishes, there is a boundary condition in the 
$SU(2)$ parent theory that remains intact. From this surviving 
boundary condition for D-branes in the $SU(2)/U(1)$ coset model, 
we have shown that the shape
of an A-brane is a straight line. 

This observation was supported by scattering amplitudes. 
In the parafermion theory
there is a global $U(1)$ symmetry, but
the consistency of the gauged WZW model for open strings
does not demand that the $U(1)$ parameter be quantized. 
Matching the results of the  $\si$-model and CFT, the scattering amplitudes
between boundary states for A-branes and closed
string states indicate that the $U(1)$ symmetry has to be
broken to  a $\Z_{2k}$ symmetry, i.e., with $2k$ end-points
on the disk boundary. The group-theoretical selection rule eliminates
half of the interval endpoints, and we are left with a 
$\Z_k$ symmetry. 

Using this $\Z_k$ symmetry, we have constructed the $\Z_k$ orbifold so that
we can implement T-duality. Orbifolding and then T-dualizing 
maps 
the original conformally flat disk to another 
conformally flat disk. The gauge field on B-branes 
was obtained from the boundary condition for A-branes.
From the resulting  $\si$-model, we derived the boundary
condition for B-branes, showing that they 
are centered disks. The obtained gauge
field strengths stabilize the B-branes, i.e.,
they prevent B-brane decay by shrinking or
displacing off the center. As the $\si$-model action 
for B-branes can be recast into the axially gauged WZW action, we conclude 
that an axial-vector duality exists in the $SU(2)/U(1)$ coset model
for open strings.

Finally, let us comment on certain differences between our results
and those of \cite{MMS}.  There 
the shape of A-branes was discussed mainly from the CFT 
point of view. Target space  
arguments were based on consideration 
of $\sigma$-models without boundaries only. In addition, while 
the field theory 
realization of the $SU(2)/U(1)$ coset model is 
the vectorially gauged WZW model, 
the axial coset was used 
to describe the geometry of its branes. 
Consequently, 
the shape of an A-brane was conjectured in \cite{MMS} to be a 
straight line in
the disk $(\rho,\phi)$. In our case, we started with the vectorially
gauged WZW action for open string.  From 
the surviving boundary condition (\ref{bc2}) and the 
$\Z_{k}$ symmetry, we showed explicitly 
that the shape of an A-brane in the $SU(2)/U(1)$
coset model is a straight line, obtained by projecting
the spherical D2-branes in the $SU(2)$
parent theory to the disk $(\rho',\tilde{\phi})$. 

Also, the authors of \cite{MMS} found that when the 
gauge field strength $F$ on 
B-branes is put in by hand, the B-branes will be stable. 
A fixed flux of $F$ on the B-brane was 
assumed and its value was calculated by 
minimizing the Dirac-Born-Infeld action. In our 
approach, however, such minimization is unnecessary. 
The gauge field strength $F$ on the B-brane is obtained more directly -- 
from the gauged WZW model, by first gauging away the vector $U(1)$ subgroup,
then orbifolding and transforming to the T-dual theory. 
The quantized gauge
field strengths in the original spherical D2-branes of the $SU(2)$ 
parent theory are inherited by these disk B-branes, making 
them stable.

\section*{Acknowledgement} 

We thank Gor Sarkissian for his 
helpful correspondence. This work was supported in part by NSERC.

\newcommand{\bib}{\bibitem}
\newcommand{\NP}[1]{Nucl.\ Phys.\ {\bf #1}}
\newcommand{\AP}[1]{Ann.\ Phys.\ {\bf #1}}
\newcommand{\PL}[1]{Phys.\ Lett.\ {\bf #1}}
\newcommand{\CQG}[1]{Class. Quant. Gravity {\bf #1}}
\newcommand{\CMP}[1]{Comm.\ Math.\ Phys.\ {\bf #1}}
\newcommand{\PR}[1]{Phys.\ Rev.\ {\bf #1}}
\newcommand{\PRL}[1]{Phys.\ Rev.\ Lett.\ {\bf #1}}
\newcommand{\PRE}[1]{Phys.\ Rep.\ {\bf #1}}
\newcommand{\PTP}[1]{Prog.\ Theor.\ Phys.\ {\bf #1}}
\newcommand{\PTPS}[1]{Prog.\ Theor.\ Phys.\ Suppl.\ {\bf #1}}
\newcommand{\MPL}[1]{Mod.\ Phys.\ Lett.\ {\bf #1}}
\newcommand{\IJMP}[1]{Int.\ Jour.\ Mod.\ Phys.\ {\bf #1}}
\newcommand{\JHEP}[1]{J.\ High\ Energy\ Phys.\ {\bf #1}}

\end{document}